\documentclass[12pt,preprint]{aastex}
\usepackage{graphicx}

\def \arcmin {\hbox{$^\prime$}}
\def \arcsec {\hbox{$^{\prime\prime}$}}
\def \oneight {\hbox{RX\, J1856.5-3754}}
\def \zeroseven {\hbox{RX\, J0720.4-3125}}
\def \zerofour {\hbox{RX\, J0420.0-5022}}
\def \zeroeight {\hbox{RX\, J0806.4-4123}}
\def \onethree{\hbox{RX\, J1308.6+2127}}
\def \onesix{\hbox{RX\, J1605.3+3249}}
\def \seventeen{\hbox{1RXS\, J214303.7+065419}}
\def \xmm{{\it XMM-Newton \/}}
\begin{document}

%----------------------------------

\title{\xmm  Detection of Pulsations and a Spectral Feature in the
X-ray Emission of the Isolated Neutron Star 
1RXS\, J214303.7+065419/RBS~1774}

\author{S. Zane\altaffilmark{1}, M. Cropper\altaffilmark{1}, R.
Turolla\altaffilmark{2}, L. Zampieri\altaffilmark{3},
M. Chieregato\altaffilmark{4} \altaffilmark{5}, J.J. 
Drake\altaffilmark{6}, A.
Treves\altaffilmark{4}}

\altaffiltext{1}{Mullard Space Science Laboratory, University College
London,
Holmbury St. Mary, Dorking, Surrey, RH5 6NT, UK; sz@mssl.ucl.ac.uk}
\altaffiltext{2}{Department of Physics, University of
Padova, via Marzolo 8, 35131 Padova, Italy; turolla@pd.infn.it}
\altaffiltext{3}{INAF-Osservatorio Astronomico di Padova, vicolo 
dell'Osservatorio 5,
35122, Padova, Italy; zampieri@pd.astro.it}
\altaffiltext{4}{Department of Physics and Mathematics, Universit\'a
dell'Insubria Via Valleggio 11, 22100, Como, Italy;
treves@mib.infn.it; matteo.chieregato@pd.infn.it}
\altaffiltext{5}{Institut f\"{u}r Theoretische Physik der Universit\"{a}t
Z\"{u}rich, Winterthurerstrasse 190, CH-8057 Z\"{u}rich, Switzerland}
\altaffiltext{6}{Smithsonian Astrophysical Observatory, MS 3, 60 Garden
Street, Cambridge, MA 02138, USA; jdrake@head-cfa.harvard.edu}

%%%%%%%%%%%%%%%%%%%%%%%%%%%%%%%%%%%%%%%%%%%%%%%%%%%%%%%%%%%%%%%%%%

\begin{abstract}

We report on the results of a  deep \xmm observation of RBS~1774, 
the most recent dim isolated neutron star candidate found
in the {\it ROSAT\/} archive data. Spectral and timing
analysis of the high-quality PN and MOS data confirm the
association of this source with an isolated neutron star. The
spectrum is thermal and blackbody-like, and there is evidence at a
significance level $> 4 \sigma$ that the source is an X-ray pulsar, with 
spin period of
9.437~s. Spectral fitting reveils the presence of an absorption
feature at $\sim 0.7$~keV, but at this level data do not have
enough resolution to allow us to discriminate between an absorption line 
or an edge. We compare the newly measured
properties of RBS~1774 with those of other known dim isolated
neutron stars, and discuss possible interpretations for the
absorption feature.

\end{abstract}

%%%%%%%%%%%%%%%%%%%%%%%%%%%%%%%%%%%%%%%%%%%%%%%%%%%%%%%%%%%%%%%%%%

\section{Introduction}
\label{intro}

Over the last decade {\it ROSAT\/} observations have led to the
discovery of seven very soft X-ray sources with quite particular
characteristics. The extreme values of the X ray-to-optical flux ratio
($\gtrsim 10^4$),
the low hydrogen column densities ($n_H\approx 10^{20}\, {\rm cm}^{-2}$) 
and the detection of
pulsations in the range $\sim 5-10$~s in four objects (five with 
the present one) strongly favor the possibility that these sources are
close-by, X-ray emitting, dim, isolated neutron stars (XDINSs; see e.g.
\citealt{t2000} and \citealt{frank04}
for reviews and Table~\ref{tab0} for a summary).

XDINSs stand apart with respect to
other known classes of bona fide isolated neutron stars detected at X-ray
energies. All of them are radio-silent, they are not associated
with supernova remnants and exhibit a soft X-ray spectrum without
evidence of a power-law hard tail.
The origin of their X-ray emission has not been
fully
clarified as yet. The recent measurement of relatively large proper 
motions
in three sources, implying transverse velocities  $\gtrsim 150\, {\rm km
\, s}^{-1}$
(\citealt{kvka2002}; \citealt{motch2003}; \citealt{motch2005}), makes
it unlikely that these are old ($\gtrsim 10^7$~yr) neutron stars
accreting from the interstellar medium (see e.g. \citealt{t2000}). Most probably the 
X-rays arise from the cooling of younger objects with
an inferred age of $\approx 10^5$--$10^6$~yr.

XDINSs play a key role in compact objects astrophysics being the
only sources in which we can have a clean view of the compact star
surface, without contamination from magnetospheric emission or 
emission from a binary
companion or a supernova remnant. They appear to be truly
``isolated'' neutron stars, and only a small number of them has been 
detected so far. Detailed multiwavelength studies of the largest possible 
sample of XDINS candidates are therefore fundamental for tracking the  
evolutionary history of galactic neutron stars, and for shedding light on 
their thermal and magnetic surface properties.

Spectral analysis carried out so far has convincingly shown that
the broadband X-ray emission from XDINSs is well represented by a
blackbody continuum, challenging the predictions of conventional
atmospheric models. The absence of spectral lines appears to be
well established in the brightest XDINS,  \oneight \
(\citealt{dra2002}; \citealt{bur2003}), while absorption features
have been increasingly reported in the spectra of other sources,
including \onethree, \zeroseven, \onesix, \zeroeight, \zerofour \
(\citealt{hab2003}; \citealt{haberl2004b}; \citealt{vk2004};
\citealt{haberl2004a}). The absorption features are quite broad, in
contrast with the atomic spectral lines predicted by atmospheric models, 
and they appear at 
energies
$\sim 200-500$~eV. Their nature is still uncertain but the
intriguing identification with a proton cyclotron
resonance has been suggested. Should this interpretation prove
correct, it makes XDINSs highly magnetized objects with fields of the 
order of, or exceeding, a few $\times 10^{13}$~G. In this
respect, the case of \zeroseven \ is particularly interesting: the
magnetic field strength derived by line energy, assuming it is a
proton cyclotron feature, is in good agreement with that implied
by the spindown measure ($\dot P\sim 3-6\times 10^{-14}\, {\rm s
\, s}^{-1}$; \citealt{silvia2002}; \citealt{kaplan2002};
\citealt{cro04}). 

We are still at a stage where every newly discovered XDINS holds important 
information for understanding the
properties of the whole class and may also show some peculiarities which 
import unique information. RBS~1774 (1RXS~J214303.7+065419) has 
been the most recent XDINSs to be found  \citep{zam2001}. The
source is listed in the {\it ROSAT\/} Bright Source catalog, and
lies about 48\arcmin \ off-axis in a PSPC pointing of the BL Lac 
MSS~2143.4+0704. The {\it ROSAT\/} observation had a limited
statistics ($\sim 500$ net counts), but it had been sufficient to
reveal a very soft X-ray spectrum, to which an absorbed blackbody
provides an acceptable fit ($kT\sim 92\, {\rm eV}$, $n_H\sim
4.6\times 10^{20}\, {\rm cm}^{-2}$). No modulation with amplitude
$\gtrsim 30\%$ has been found in the PSPC data. Optical follow-up
observations have shown no plausible optical counterpart in the
X-ray error box down to $R\sim 23$, implying an X-to-optical flux
ratio $\gtrsim 10^3$. All these properties are very similar to
those of the already established XDINSs and make RBS~1774 worth of
further investigation.

In this paper we report results from a recent \xmm
observation of RBS~1774 and compare the newly measured properties
of this source with those of other known members of this class.

%%%%%%%%%%%%%%%%%%%%%%%%%%%%%%%%%%%%%%%%%%%%%%%%%%%%%%%%%%%%%%%%%%
\section{X-ray Observations}
\label{obs}

RBS~1774 was observed with \xmm on 2004 May 31 for 30~ks (EPIC
PN). All the three EPIC detectors were configured in small window 
mode. We processed the data using the \xmm SASv6.0.0. We selected
periods of low background, with the Good Time Interval (GTI) file
leaving an effective exposure of 23~ks (EPIC PN). The same GTI
was used for the spectral and timing analysis.

\begin{figure}
\includegraphics[width=5in,angle=0]{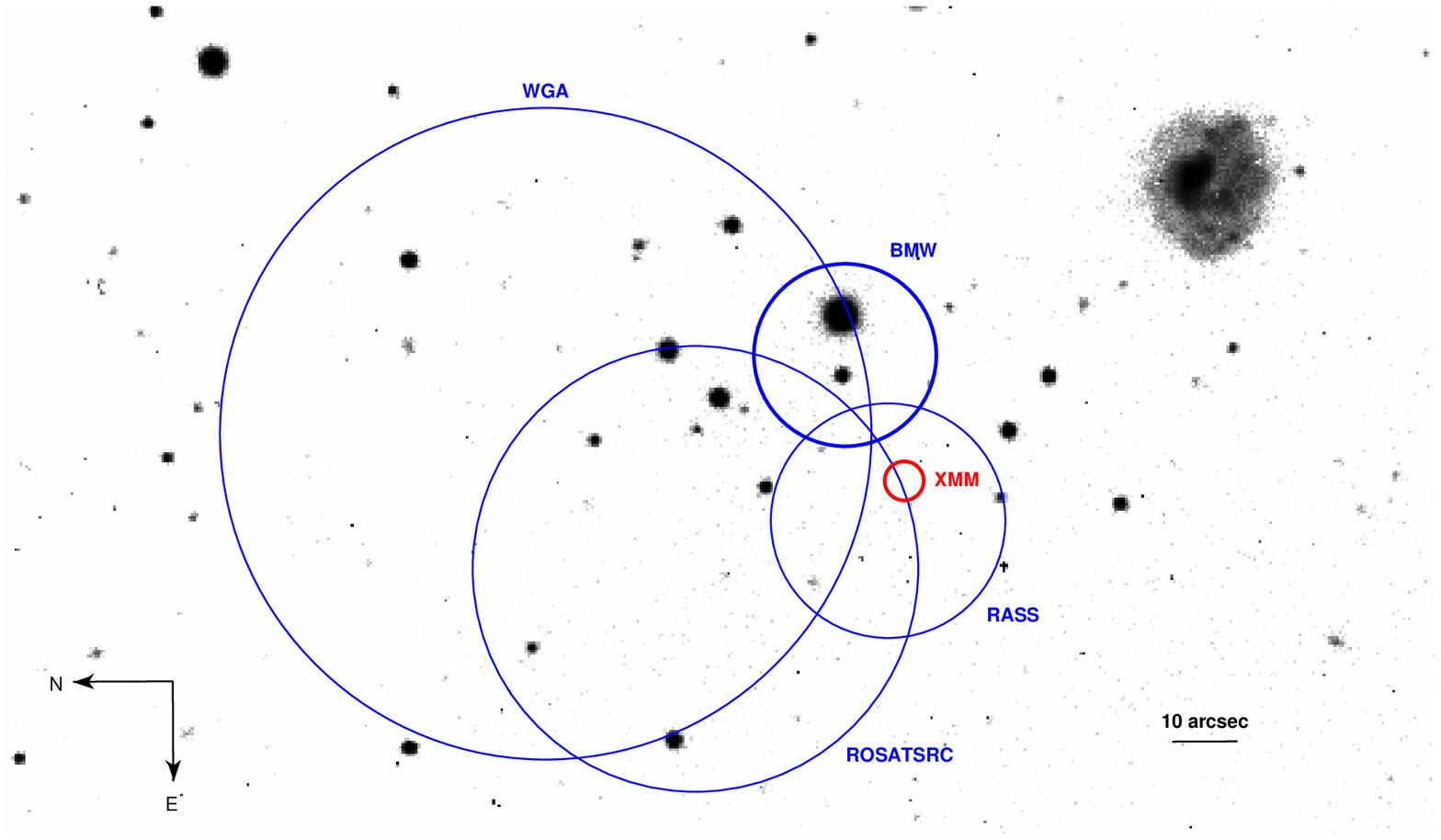}
\caption{\label{pos}
R-band NTT image of the field around RBS~1774 (limiting magnitude 
$\sim 22.8$, see text and 
\citealt{zam2001}). The red circle (3" radius) shows the source position
as derived from our recent \xmm \ observation, while the blue circles 
show the position inferred from past data (see the electronic version 
for a full color image). Positions (see \citealt{zam2001} for all details) 
are  taken from: 
\cite{white94} 
Catalogue (WGA);  
{\it ROSAT\/} ASS Bright Sources Catalogue (RASS, \citealt{voges96}), 
{\it ROSAT\/}  
SRC Catalogue (ROSATSRC, \citealt{zimme94}), RASS field analized with a 
Brera 
Multiscale Wavelet alghoritm (BMW, \citealt{lazz99}). 
Error circles represent the  90\% confidence 
intervals.}
\end{figure}

The source is clearly visible in EPIC PN, MOS1 and MOS2 with count
rates of $2.054 \pm 0.011$, $0.346 \pm 0.004$ and $0.328 \pm
0.004 $ counts s$^{-1}$ respectively (0.2-2~keV band).
%\footnote{The total 
%counts detected by the RGS instruments were not
%sufficient for a meaningful spectral analysis.} 
In the EPIC image, the source is slightly off axis since the pointing was 
requested based on the revised position published by
\cite{zam2001}. The position obtained with \xmm (J2000 coordinates) is 
$\alpha = 21^h 43^m
03.3^s$, $\delta = +06^\circ 54'17'' $, with a $90\%$ uncertainty radius
of $3\arcsec$. Within 3$\sigma$, previous estimates based on {\it ROSAT\/}
data (see again \citealt{zam2001}) are consistent with the XMM position.
In particular, the position from the {\it ROSAT\/} ASS Bright Sources
Catalogue ($\alpha = 21^h 43^m 03.7^s$, $\delta = +06^\circ 54'19.5'' $,  
90\% uncertainty radius of $18\arcsec$; labelled with RASS in
\citealt{zam2001}) is within $\sim 7 \arcsec$ from the present XMM
position. Figure~\ref{pos} shows the {\it ROSAT\/} and \xmm error boxes
overlayed on the optical image (discussed below). We note the small
difference in the position of the RASS error box with respect to that   
reported by \cite{zam2001}. 

\subsection{Spectral analysis}
\label{spec}

For the EPIC PN, we extracted single and double events ({\sc pattern}
$\leq 4$, FLAG=0) within a circle of 25\arcsec \ radius for source and 
within
two circular regions (to avoid out-of-time events) with radius 50\arcsec \  
for background. For the EPIC MOS1 and 2, we extracted up to
quadruple events ({\sc pattern} $\leq 12$, FLAG=0) with a 25\arcsec \ and 
32\arcsec \ 
radius for the source respectively with two rectangular background
regions to exploit the limited window size more effectively. In
order to account for the slightly off-axis position, we generated
the appropriate off-axis response and ancillary response
matrices. Data were grouped, ensuring a minimum of 30 counts per
energy bin.

\begin{figure}
\includegraphics[width=5in,angle=0]{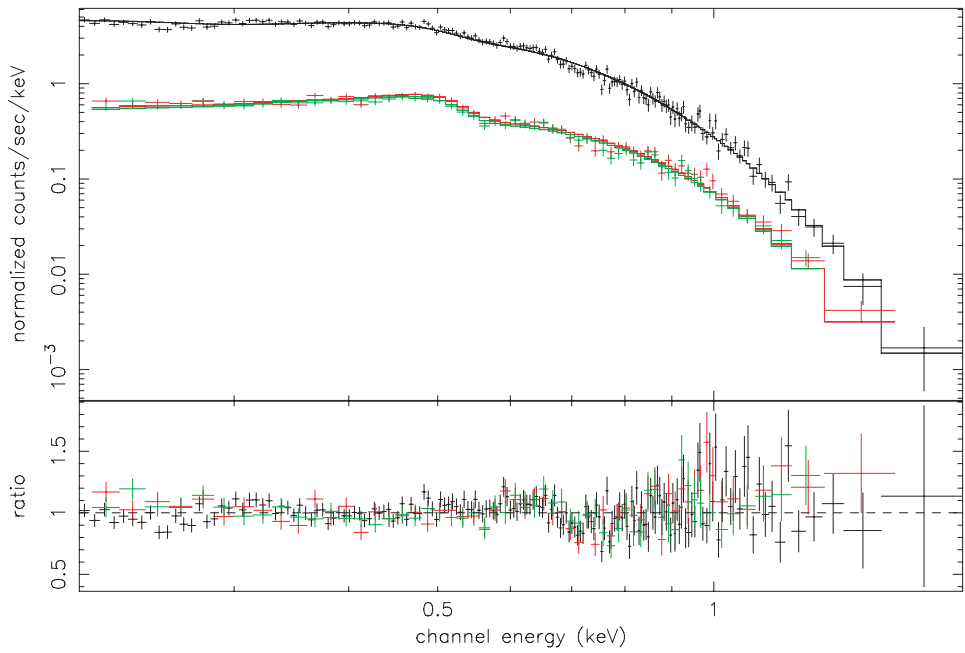}
\caption{\label{bbfit} The top panel shows the count rate spectrum
of RBS~1774 obtained with EPIC PN (black) and the two EPIC MOS
(light green, red in the color version) detectors, with the best single 
blackbody combined
model fit (parameters are given in Table~\ref{tab1}). The bottom
panel shows the data to model ratio.}
\end{figure}

We performed a spectral analysis by fitting data from the three
EPIC detectors simultaneously. We find that the broadband spectrum
can be represented by a single absorbed blackbody, with $kT
=0.1014$~keV and $n_H = 3.65
\times 10^{20}$~cm$^{-2}$ (see Figure~\ref{bbfit},
Table~\ref{tab1}). Both parameters are consistent with
those measured in the past using {\it ROSAT} data, although better
constrained. The unabsorbed EPIC-PN flux in the $0.1-2.4$ keV band is
$6.05\times10^{-12}$ erg cm$^{-2}$ s$^{-1}$, slightly less than that
reported by Zampieri et al. for {\it ROSAT} PSPC ($8.7\times10^{-12}$ erg
cm$^{-2}$ s$^{-1}$ (the difference is likely to be due to the large 
off-axis angle of RBS~1774 in the {\it ROSAT} observation). Assuming a 
typical  
luminosity of $5\times 10^{31}$~erg/s gives a distance $\sim 280$~pc. The 
value of the column 
density is similar to, but slightly 
lower than, the total Galactic absorption in the source direction ($n_H = 
5 \times 10^{20}$~cm$^{-2}$, \citealt{dl90}). The resulting reduced 
$\chi^2=1.36$ 
is not fully satisfactory and an inspection of the residuals suggests that 
the
largest discrepancies between model and data are above $\sim 0.6$~keV.

As discussed by many authors (see e.g. \citealt{zp2002}, 
\citealt{t2000} for reviews), 
realistic model spectra of the cooling
atmosphere surrounding an unmagnetized ($B\lesssim 10^9$~G)
neutron star exhibit a distinctive hardening with respect to a
blackbody and deviate significantly from a Planckian shape. When
accounting for a magnetic field of moderate strength ($B\sim
10^{12}-10^{13}$~G), the hard tail present in nonmagnetic models
with comparable luminosity is partially suppressed and the X-ray
spectrum, although still harder than a blackbody at the neutron
star effective temperature, $T_{eff}$, is more Planckian in shape.

Models of fully ionized, pure H neutron star cooling atmospheres,
as computed by \cite{p1992} and \cite{z1996}, are currently
implemented in XSPEC ({\it nsa \/} model) for three different values of 
the
magnetic field: $B=0$ (unmagnetized models) and $B=10^{12},
10^{13}$~G. In the attempt to improve the spectral fit we tested
all the three sets of models, but in all cases the resulting
$\chi^2$ was worse than that obtained using a simple blackbody
model (see Table~\ref{tab2}). The star mass and radius have been
fixed at $M=1.4M_\odot$ and $R=10$~km, but the effects of
gravity on the emergent spectra are too small to produce any
appreciable difference in the fits.

\begin{figure}
\includegraphics[width=5in,angle=0]{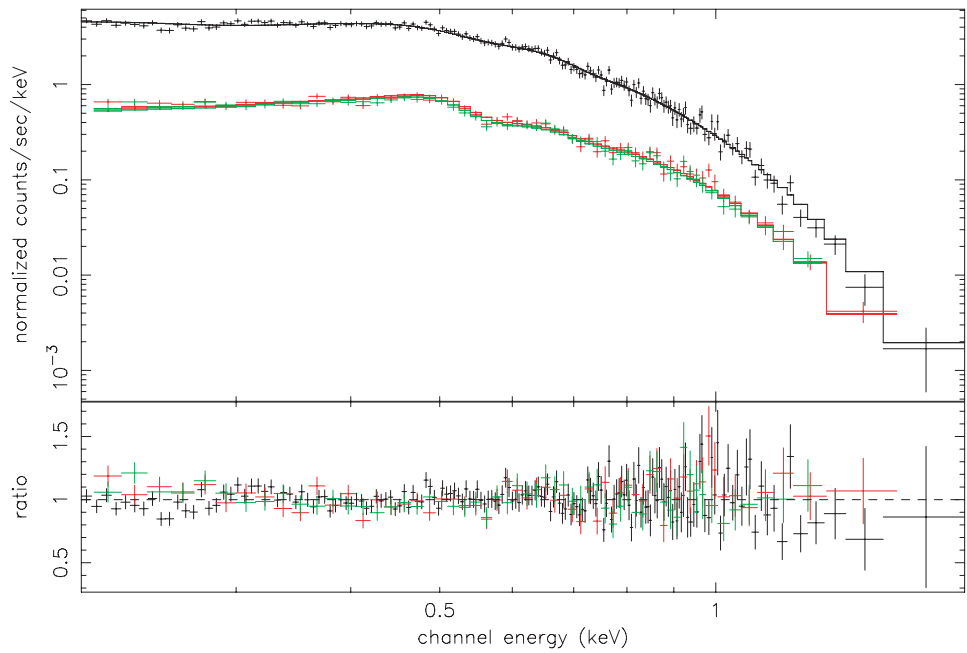}
\caption{\label{bbedgefit}
Top panel: the count rate spectrum of RBS~1774 obtained with EPIC
PN and the two EPIC MOS detectors. The best fit model is an absorbed
blackbody with an absorption edge at $\sim 0.7$~keV (parameters are given
in Table~\ref{tab1}). The bottom panel shows the data to model ratio.}
\end{figure}

Similarly, no improvement is obtained adding a power-law
component or a second blackbody. 
However, we find that adding a gaussian 
line in
absorption at $\sim 0.7$~keV produces a statistically significant 
($\Delta\chi^2=56$; $7\sigma$) improvement\footnote{The F-test has been 
widely used to test the
significance of spectral lines, although this is strictly
inappropriate (see \citealt{prot2002}).  Only for completeness, we
report that we have checked the F-test statistics value and
probability, obtaining 12.1 and $1.8 \times 10^{-7}$,
respectively.}, leading to a reduced $\chi^2 = 1.20$. The best fitting
energy of the line is $E_{line}= 0.754$~keV; the
line width and depth are $\sigma_{line}
=0.027 $~keV and $\tau_{line} =
4.8$, respectively. The latter two parameters are mutually degenerate and 
not well constrained, because, at least for narrow or moderately narrow lines,
the energy resolution is insufficient to strongly distinguish
between width and depth.
%It should be noted that a narrow line is hugely
%broadened by the spectral response at these energies.
A slightly better (reduced $\chi^2 = 1.17$) but similar fit is obtained by
using an absorption edge instead of a gaussian line (see 
Figure~\ref{bbedgefit}): in the edge
model there is one parameter less, therefore the depth is well
constrained. At this level the data do not have enough resolution to
allow us to discriminate between the two kinds of spectral
features, in terms of their physical meaning. The best fitting
energy of the edge is $E_{edge}=0.694$~keV and
the optical depth is $\tau_{edge} = 0.25$. In both
cases, the blackbody temperature is essentially unchanged with
respect to the single blackbody fit. The best fit parameters are
reported in Table~\ref{tab1}. The flux in the line/edge has been 
computed by 
fitting data with the models above, then switching the optical depth to 
zero and calculating the difference in flux, yielding $8.6 \times 
10^{-14}$~erg~cm$^{-2}$s$^{-1}$ (edge) and $3.3 \times
10^{-14}$~erg~cm$^{-2}$s$^{-1}$ (gaussian line).

\subsection{Timing analysis}
\label{tim}

The high throughput of \xmm provides a much more stringent check
than the {\it ROSAT} PSPC on whether RBS~1774 is pulsating. For the timing 
analysis, we extracted the counts within a 30\arcsec 
\ radius 
aperture in both PN and MOS1 and MOS2 but with the
aperture truncated by a chord where it approached the edge of the
CCD window in each case: this maximised the counts while avoiding
any edge effects.  A background area was selected using as much as
possible of the small window mode image area while avoiding the
edge of the window and the out-of-time events. This resulted in
event files containing 38485, 9738 and 9500 events available for
data analysis for PN, MOS1 and MOS2 respectively.

We analysed PN, MOS1 and MOS2 data both individually and combined
using the Maximum Likelihood Periodogram technique described in
\cite{silvia2002} and \cite{cro04}. We searched periods from 10000~s to 
30~ms,
ensuring that in each case the period grid was 2.5 times better
sampled than the Nyquist frequency. Below 0.3~s the time
resolution even in small window mode of MOS prevents them
being used, so only the PN data were used for periods shorter than
this.

\begin{figure}
\includegraphics[width=5in,angle=0]{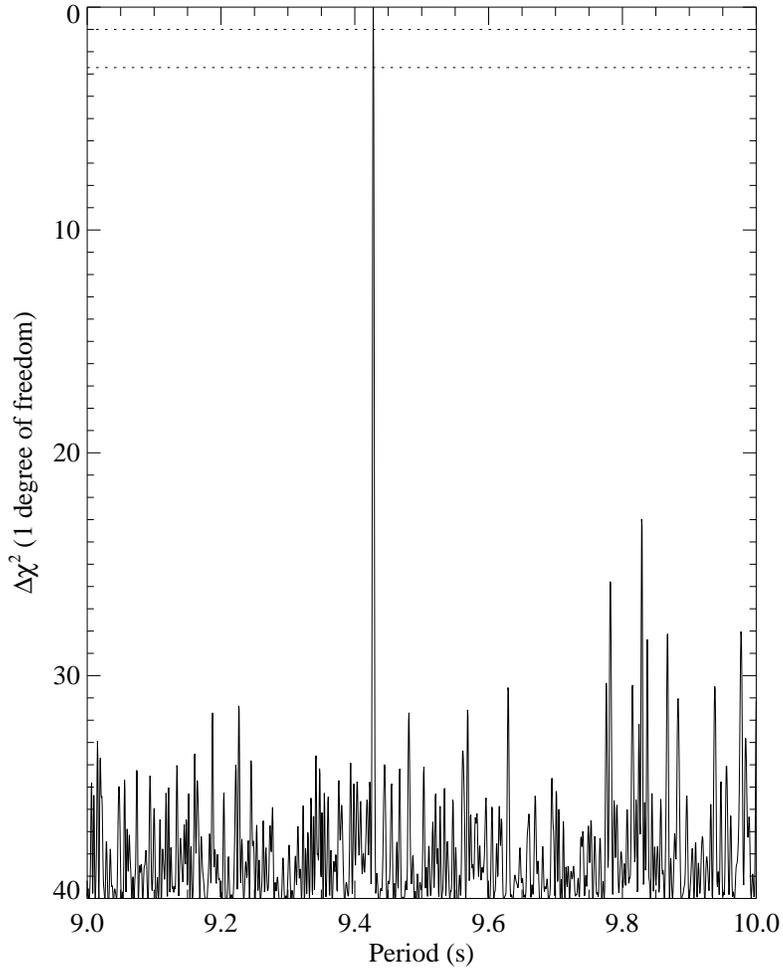}
\caption{\label{periodog}
Maximum Likelihood Periodogramme from the combined
EPIC MOS and PN data of RBS~1774 (see text for details).}
\end{figure}

The search revealed a significant period at 9.437~s. This is illustrated
for the combined MOS and PN dataset in Figure~\ref{periodog}, where the 
dotted 
lines show confidence intervals of 68\% and 90\%. The $\Delta\chi^2$
between the maximum and the noise floor is $40$ in the combined dataset,
while for the PN alone $\Delta\chi^2 = 36$. The peak is evident in the
combined MOS1 and MOS2 data, with $\Delta\chi^2 = 9.5$. We have checked
whether the reduced significance of the MOS1+2 peak is consistent with
that in the PN data, given its greater sensitivity, by selecting (at
random) event data from the PN event list to generate a reduced dataset
with the same number of events as MOS1+2. In this reduced PN dataset the
noise levels are similar to the MOS1+2 dataset, and the 9.437~s peak is no
longer the most significant in the 9--10 sec period range. This indicates
that the lower significance of the peak in the MOS1+2 dataset is 
consistent with its reduced statistics with respect to the
PN dataset and that no additional bias is introduced because
of event selection.

$\Delta\chi^2 = 40$ corresponds to a $>6 \sigma$ result 
($\sigma=\sqrt{\Delta\chi^2}$ for 1 degree of freedom). However,
we need to consider the random probability of a peak occurring
given the large number of periods searched. 
In the interval between 1 and 1000
seconds, there are 59638 independent periods, and the probability
of a $6 \sigma$ peak will occur once in $1.5\times10^5$ cases,
corresponding to a $\Delta\chi^2=18.7$ or $>4\sigma$ result. We
therefore consider the peak to be significant, although with the caveat 
that 
in \cite{hpm99} a $4\sigma$ detection was claimed for 
a 22.69~s
periodicity in  \zerofour, which has since been shown to have a 3.45~s
periodicity.

\begin{figure}
\includegraphics[width=5in,angle=0]{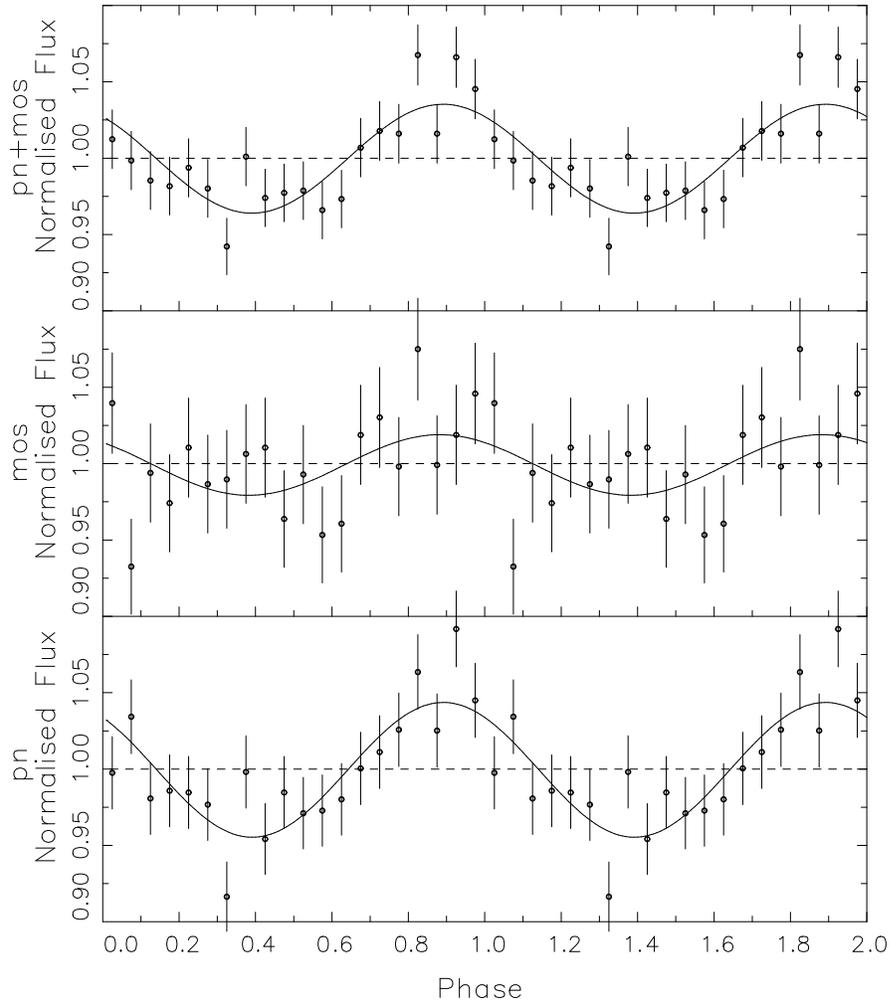}
\caption{\label{lc}
Folded 0.15-3~keV EPIC PN and MOS light curve of 
RBS~1774.}
\end{figure}

We have consequently folded the data on the 9.437~s period and
binned them in phase. This is shown in Figure~\ref{lc} for PN
separately, for the combined MOS data, and then for all three
instruments together. The MOS data taken separately are noisy, with a
variation which would not be considered significant in its own
right, but which nevertheless shows an amplitude and phasing
which is consistent with that from the PN. This is evident (equivalently) 
from our periodogram analysis where the
$\Delta\chi^2$ increases from 36 to 40 when the MOS1+2 data are added to 
the PN data, and adds weight to our conclusion that this pulsation is 
real. 

The pulsation is approximately sinusoidal, and the amplitude of
the variation is remarkably small: fitting a sinusoid yields a
semiamplitude of $0.036\pm 0.006$ (0.072 peak-to-peak). In the
attempt to confirm the periodicity, we re-analysed the archival
{\it ROSAT\/} data taken in May 1991 (\citealt{zam2001}). This
observation is divided into two segments of duration  $\sim
1800$~s each; the interval between the two segments is $\sim
4000$~s. No statistically significant peak is present in the power
density spectrum of the two separate segments, due to the poor
statistics. No modulation with amplitude
$\gtrsim 30\%$ has been found in the PSPC data, and this limit is 
consistent with the \xmm detection.

%------------------------------------------------------------------

\section{Possible nature of the spectral feature}
\label{line}

The quality of the blackbody fit to the phase-averaged EPIC data
of RBS~1774 is not completely satisfactory. Adding an absorption
edge or a gaussian line in absorption significantly improves the
spectral fit, suggesting the presence of an absorption feature
similar to those observed in \onethree, \zeroseven, \onesix,
\zeroeight, \zerofour \ (\citealt{hab2003}; \citealt{haberl2004b};
\citealt{vk2004}; \citealt{haberl2004a}). However, the absorption
features detected in these other sources appear at energies below 
$500$~eV, while in the case of RBS~1774 it is at $\sim 700$~eV (for
the purpose of the discussion we use the energy of the feature
obtained by the blackbody plus edge fit, which is slightly better
constrained). This makes the detection in  RBS~1774 less sensitive to
the uncertainties in the EPIC-PN spectral calibration, since
residual systematic calibration effects are mainly present below
$\sim 0.5$~keV.  We note that a feature at this energy has been
detected in the spectrum of the neutron star 1E 1207.4-5209
(\citealt{sptz2002}; \citealt{mer2002}; \citealt{big2003};
\citealt{hk2002}; \citealt{m2004}; \citealt{zav2004}). The comparison 
between the properties of the XDINSs and those of 1E~1207.4-5209 must be 
done carefully, given the fact that the latter is probably younger: it 
has a shorter period, it is 
still associated with a supernova remant and it shows an additional 
harder spectral component in the X-ray spectrum. Also, 
1E~1207.4-5209 is unique since it exhibits at least two spectral
lines, with harmonic spacing. However, as will be discussed further on,
some of the models proposed for this source can be applied to
RBS~1774 as well.

Strong atomic absorption features are predicted by low-field atmospheric models with
significant metal abundances (either solar, or pure iron, see
\citealt{ra96}; \citealt{z1996}). The predictions for strongly magnetized
metal-enriched atmospheres are more difficult to assess, due to the
lack of detailed calculations. However, the strongly magnetized pure Fe spectra
presented by \cite{ra97} also show a significant
number of absorption lines. The main difficulties faced by these models
is that they exhibit a variety of rather narrow lines, in contrast with
the single feature observed in RBS~1774.

Other possible interpretations are discussed below and can be divided in
two groups, requiring either a superstrong ($B \sim 10^{14}$~G) or a
moderate ($B \sim 10^{10}-10^{12}$~G) magnetic field.

\subsection{Super-strong magnetic fields}

A plausible hypothesis is that the feature in the spectrum of
RBS~1774 is associated with a proton cyclotron line. An
interpretation in terms of a proton resonance has been recently
put forward for the lines observed in other XDINSs, and, at least
for \zeroseven, it is supported by the agreement between the
magnetic field strength independently inferred from spin-down
measurements and line energy (\citealt{cro04}). In the case of
RBS~1774, this would imply an ultra-strong magnetic field of $B
\sim (E_{edge}/0.63\, {\rm keV}) (1+z) \times 10^{14} \, {\rm G}
\sim 1.4 \times 10^{14}$~G, assuming a gravitational redshift
factor at the star surface of $(1+z)^{-1}=\sqrt{1-2GM/c^2R}\sim
0.8$. Such a field strength is comparable (although a factor $\sim
30\%$ higher) with that estimated in a similar way for \onesix \ and
\zeroeight \ ($\sim 9 \times 10^{13}$~G, see \citealt{vk2004};
\citealt{haberl2004a}), and is well above the critical value,
$B_Q \simeq 4.4 \times 10^{13}$~G, at which QED effects become important. 

An alternative interpretation is that we are detecting an
absorption feature due to a bound-bound or bound-free transition
in the hydrogen atom. A strong magnetic field can significantly
increase the atomic binding energies and radiative transition
rates, moving the H ionization edge into the soft X-ray band. First
calculations of hydrogen atmospheres with bound atoms have shown
that broad spectral lines become prominent if the atmospheric
temperature drops below $\sim 10^6$~K. The strongest line
corresponds to the transition between the ground state and the
lowest excited state at an energy of $\approx
75[1+0.13\ln(B/10^{13}~{\rm G})] + 63 (B/10^{13}~{\rm G})$~eV
(\citealt{zp2002}). If applied to the $\sim 0.7$~keV line detected in 
RBS~1774, this interpretation yields a magnetic field of $B \approx
10^{14}$~G, again in the ultra-magnetized range. More detailed atmospheric
models which include partially ionized hydrogen have been computed
by \cite{ho03}, accounting for magnetic field effects and using
the latest available equation of state. They have shown that, for
$B=5 \times 10^{14}$~G and $T=5 \times 10^6$~K, two broad
absorption features, due to bound-free transitions to different
continuum states, are present at $\sim 0.76$ and 4~keV,
respectively. The latter is much stronger, but it would not be
visible in the spectrum of RBS~1774 which has negligible emission
above $\sim 2$~keV. At somewhat lower fields ($B\sim 10^{14}$~G) 
\cite{ho04} found that 
the dominant atomic features in the 0.2--2~keV band are those due
to the low-energy bound-free transition, which is barely visible at 
0.54~keV,
and the bound-bound transition discussed above at 0.74~keV. 
These authors argue that at super-strong
field strengths ($B\gtrsim 10^{14}$~G) vacuum polarization may
efficiently suppress these lines, in which case the above conclusions can 
be (at least partially) invalidated. However, current treatment of this
QED effect is too crude to make a definite statement.

At present, accurate atmospheric models for chemical compositions other 
than hydrogen are unavailable for  $B\gtrsim 10^{12}$~G. In
the case of 1E 1207.4-5209, attempts to interpret the $\sim
0.7$~keV feature as a transition in ions other than H have
been made, simply based on the latest theoretical results on the 
characteristics of bound-bound and
bound-free transitions in He-like and H-like ions in  strong
fields. In this case, a $\sim 0.7$~keV feature may be associated with a 
transition  between the ground level ($m=0, \nu=0$, where $m$
and $\nu$ are the magnetic and longitudinal quantum numbers) and
the lowest excited tightly-bound level $(m=0,\nu=1)$ of ${\rm
He}^+$. The required magnetic field is $\sim 1.4 \times
10^{14}$~G, for $(1+z)^{-1} \sim 0.81$ (\citealt{sptz2002};
\citealt{zav2004}). Taken overall, these results show that if
RBS~1774 has a magnetar-like field, $B\approx 10^{14}$~G,
the feature at $\sim 0.7$~keV may be due to a proton cyclotron
resonance, or bound-bound transitions in H or H-like He, or a blend of 
these. 

\subsection{Moderate magnetic fields}

An obvious scenario involving a moderate magnetic field is that in
which the observed feature is associated with  electron (instead of
proton) cyclotron resonance. Would this be the case, the surface
magnetic field is necessarily much lower, $B\sim  (E_{edge}/11.6\,
{\rm keV}) (1+z) \times 10^{12} \, {\rm G} \sim 7.5 \times 10^{10}
\, {\rm G}$ for $(1+z)^{-1} \sim 0.8$.

A further possibility is that of an atomic transition out of the
$(m=0,\, \nu=0)$ ground state to the $(m=0,\, \nu=1)$ excited
state or to the continuum in a mid-$Z$ H-like ion, such as C~VI,
N~VII or O~VIII. Such an interpretation has been put forward by
\cite{hk2002} again in connection with the lines in 1E 1207.4-5209
and yields $B\sim 10^{11}$~G (we refer in particular to their
solutions B; see also their discussion about the possibility of
having elements other than H in the neutron star atmosphere).
Both these models are conceivable, although the required field
strength is well below that estimated for \zeroseven \ and well
below the average fields measured in ordinary radio pulsars (only
$\sim 20\%$ of the pulsar population has $B \leq 6 \times
10^{11}$~G).

\section{Comparison with other thermally emitting Neutron Stars and open
issues}
\label{discuss}

In this paper we have reported on the results of a recent \xmm
observation of RBS~1774. Spectral and timing analysis of the
high-quality PN and MOS data confirm the association of this
source with an isolated neutron star (as originally reported by 
\citealt{zam2001}). The spectrum is thermal and
there is evidence at a
significance level $>4 \sigma$ that the source is an 
X-ray pulsar, with spin period of 9.437~s. The EPIC-PN X-ray light curve
is single peaked and approximately sinusoidal, in agreement with
what is seen in most of the other XDINSs (see \citealt{cro2001};
\citealt{haberl2004a}; \onethree \ is the only XDINSs with double peaked 
X-ray pulse). However, both the relatively low level of
significance of the periodicity and the small amplitude variation
in the folded data prevented us from performing pulsed phase
spectroscopy. 

The deepest available optical pointing of the field
is that obtained with the New Technology Telescope (NTT) at la
Silla on 2001 May 27. The faintest source in this image is object
D, as listed in Table~2 by \cite{zam2001}, with $R=22.77$. This,
combined with the unabsorbed \xmm X-ray flux ($f_X = 
5 \times 10 ^{-12}$~erg\ cm$^{-2}$s$^{-1}$ in the 0.2-2 keV band for $n_H 
= 
3.6 \times 10^{-20}$~cm$^{-2}$) allows a 
lower limit 
to be placed on the X-to-optical flux ratio of $f_X/f_R \sim
1700$. When 
using the absorbed flux 
$f_X = 2.8 \times 10^{-12}$~erg \ cm$^{-2}$s$^{-1}$ and the same X-ray 
energy 
band, we get $f_X/f_R \sim 1000$.  

Taken overall, the characteristics of RBS~1774 are remarkably
similar to those of an other XDINS recently studied in detail,
\zeroeight \ (see \citealt{haberl2004a}). Both sources have a
similar EPIC-PN count rate of $\sim 1.7-2.0$ counts 
s$^{-1}$ and a similar period ($\sim 10$~s).\footnote{\onethree \   
also has a similar count rate, but a much higher 
pulsed fraction.}
Also, they have a blackbody
temperature of $\sim 100$~eV, among the highest observed from the
XDINs, and exhibit the smallest pulsed fractions of the group
(semiamplitude of $4\%$ and $6\%$ respectively) which makes the
detection of significant spectral variations with phase difficult.
Intriguingly, and if related, the two latter characteristics may
be suggestive of a nearly pole-on viewing geometry, with the spin and 
magnetic axes fairly closely aligned. This appears
also broadly consistent with the fact that the energy of the
spectral feature is the highest in RBS~1774,  \zeroeight \ and
\onesix. If the feature energy is sensitive to the field strength, then,   
both the temperature and the average magnetic field are
higher in the polar regions.

As discussed in \S~\ref{line}, many of the viable interpretations
for the feature detected at $\sim 0.7$~keV require a magnetar-like
field strength,  $B\sim 10^{14}$~G. For a period of $9.437$~s, this 
implies a  dipole spin-down rate (at present) of $\dot P \sim 11 \times 
10^{-13}$~s/s, and a spin-down age of $1.4 \times 10^5$~yrs (assuming $B$ 
constant). 
In order to investigate 
if such a high magnetic field strength is compatible with the cooling age 
of the source,  
we considered the evolutionary tracks in the $B$-$\dot P$ diagram by 
repeating the simple computation discussed in \cite{cro04}. 
We computed the source age, $\tau_d$, and the value of the magnetic field
at the birth of
the neutron star, $B_0$, obtaining 
$B_0 \approx 
1.6 \times 
10^{15}$~G and $\tau_d 
\approx 10^4$~yrs for Hall cascade; $B_0 \approx (1.1 \ {\rm or} \ 1.5) 
\times 
10^{14}$~G and 
$\tau_d \approx (1.4 \ {\rm or} \ 1.1) \times 10^5$~yrs for ambipolar 
diffusion in the  
solenoidal or irrotational mode, respectively. 
The situation is therefore 
very similar to that of  \zeroseven \ 
(\citealt{cro04}): ambipolar diffusion predicts a magnetic field which is 
relatively
constant over the
source lifetime, while  Hall cascade gives 
a scenario compatible with a superstrong field at the star's birth ($B_0 
\sim 10^{15}$~G). The corresponding 
star's  ages are compatible with the cooling age of a neutron 
star with mass $\approx 1.35 $~M$_\odot$ or slightly larger, 
$1.4-1.5$~M$_\odot$ (see Fig.~2 in 
\citealt{ya04}). 

Should the strength of the magnetic field be confirmed by future
observations, this will make RBS~1774 the XDINS with the highest field
discovered so far. \onesix \ and \zeroeight \ may possess magnetic fields
of comparable strength ($B\sim 9 \times 10^{13}$~G \citealt{vk2004};
\citealt{haberl2004a}), and this raises the question of whether or not a
highly magnetized neutron star may remain for long periods without any
sign of some kind of transient activity.  Very recently, a few radio 
pulsars 
with $B>B_Q$ and long period ($\approx$ few seconds) have been discovered 
(see e.g. \citealt{mcl2003};  
\citealt{m2002}; \citealt{ca2000}). On the other hand, non-thermal 
magnetospheric emission from XDINSs is so far undetected, although 
this may  simply reflect the fact that it
falls below the detection threshold of present instruments. In fact, 
if XDINSs are similar to
the standard radio-pulsars the relation between rotational energy and
non-thermal X-ray luminosity would be $L_X \approx 10^{-3} \dot E$. 
Taking the above value for the magnetic field gives $\dot E \approx 
10^{31}$~erg/s, for RBS~1774, implying $L_X \approx10^{28}$~erg/s. 

XDINSs were unanimously believed to be steady
sources, as indicated by several years of observations for the
brightest of them. Only recently, and quite surprisingly, \xmm \ 
observations
revealed a substantial change in the spectral shape and pulse
profile of \zeroseven \ over a timescale of $\sim 2$~yr
(\citealt{dev2004}; \citealt{vink2004}). Possible variations in
the pulse profile of  \zerofour \ over a similar timescale ($\sim
0.5$~yr) have also been reported \citep{haberl2004a}, although at
a much lower significance level. Even more recently, timing analysis
of {\it Chandra\/} and \xmm \ data of 1E 1207.4-5209 revealed
non-monotonic variations in the spin period, possibly related to
glitches occurring on a timescale of $\sim$~1-2~yr
(\citealt{zav2004}). 
Also the monitoring of anomalous X-ray pulsars (AXPs), previously
believed to be steady emitters, has now revealed X-ray variability: both 
short 
energetic bursts and long-lasting variations in the X-ray emission of some 
of these sources have been reported 
(see \citealt{gkw03}, \citealt{woods04}, and
\citealt{sandro04} for the cases of 1E 2259+58 and 1E 1048-59).
Quite interestingly, the long-term changes observed in \zeroseven
\ bear some resemblance to those reported in 1E 2259+58 and 1E
1048-59, with the difference that in \zeroseven \ the spectral and
lightcurve evolution is not accompanied by any flux 
variation.\footnote{And again, we caveat that given the several spectral 
and timing  differences, a direct comparison between XDINSs and AXPs or 
sources like  
1E~1207.4-5209 must be done with caution.}

Further timing studies of XDINSs will be of key importance in securing new 
spin-down 
measurements: 
this will provide a second independent measurement
of the field strength and also will bring further insights on the long-term
behavior of XDINSs. Should the presence of ultra-strong fields and
the variability be confirmed, the putative relation between XDINSs
and AXPs, put forward on the basis of the similarity in spin
periods (see e.g. \citealt{sa2002}, \citealt{frank04}), becomes much 
firmer. In turn, this relation may provide the
``missing link'' among XDINSs and SGRs. If indeed XDINSs are the
descendants of SGRs/AXPs, as their lower temperature may suggest, then 
explaining why evolution produces ``quiet'' magnetars is still a
challenge.

%%%%%%%%%%%%%%%%%%%%%%%%%%%%%%%%%%%%%%%%%%%%%%%%%%%%%%%%%%%%%%%%%%%%

\begin{acknowledgements}
We thank G.G. Pavlov for a discussion about He$^+$ transitions in the 
presence of strong magnetic fields, in connection with the model presented 
by \cite{sptz2002}. RT, LZ and AT acknowledge financial support from the 
Italian Ministry for Education, University and Research through grant
PRIN-2002-027245 and PRIN-2004-023189. JJD acknowledges
support from  NASA contract NAS8-39073 to the {\em Chandra X-ray
Center. We thank an anonymous referee for several useful comments on 
the manuscript.} 

\end{acknowledgements}

%%%%%%%%%%%%%%%%%%%%%%%%%%%%%%%%%%%%%%%%%%%%%%%%%%%%%%%%%%%%%%%%%%%%

%%%%%%%%%%%%%%%%%%%%%%%%%%%%%%%%%%%%%%%%%%%%%%%%%%%%%%%%%%%%%%%%%%%%

\begin{deluxetable}{lcccccc}
\tablecolumns{4} 
\tablewidth{0pc} 
\tabletypesize{\footnotesize}
\tablecaption{Summary of XDINSs properties \label{tab0}}
\tablehead{ \colhead{Source} &  
\colhead{$kT_{bb}^\infty$ } & 
\colhead{$n_H$} & \colhead{$E_{line}$ }&  \colhead{$P$ } & 
\colhead{Semi-Ampl.\tablenotemark{a}} & \colhead{Ref.\tablenotemark{b}} \\
\colhead{ }  &  
\colhead{ eV } & 
\colhead{$10^{20}$ cm$^{-2}$} & \colhead{ eV }&  \colhead{ s } & 
\colhead{ } & }
\startdata
RX J1856-3754   &  56.7 & 0.18 &  no &  no & no & 1\\
RX J0720-3125   &  85.2 &  $1.38 $ & 270  & 8.39 & 
11\% & 2 \\
RX J1605.3+3249 &  94.1 & 0.68 & 493  & no  & no 
& 
3 \\
RX J1308.6+2127 &  85.8 & 4.10 & 290  & 10.31  & 18\% & 4 \\
RX J0420-5022   &  44.9 &  1.02 & 329  & 3.45  & 
13\% & 5 \\
RX J0806-4123  &  95.6 & 0.41 & 460  & 11.37  & 
6\% & 5 \\
RBS 1774        & 101.4 & 3.65 & 700  & 9.44  & 4\% & 6 \\ 
\enddata
\tablenotetext{a}{Semi amplitude of the folded light curve.} 
\tablenotetext{b}{1) \cite{bur2003}; 2) \cite{haberl2004b}; 3) 
\cite{vk2004}; 4)  
\cite{hab2003}; 5) 
\cite{haberl2004a}; 6) this paper.}
\end{deluxetable}

\begin{deluxetable}{lccccccc}
\tablecolumns{4} 
\tablewidth{0pc} 
\tabletypesize{\footnotesize}
\tablecaption{Model Fit
Parameters: blackbody (bb) model.\tablenotemark{a}\label{tab1}}
\tablehead{ \colhead{Model} & \colhead{$n_H$} &
\colhead{$kT_{bb}^\infty$} & \colhead{$E_{edge/line}$} &
\colhead{$\tau_{edge}$; $\tau_{line}$} & $\sigma_{line}$ &
\colhead{$f_X$\tablenotemark{b}} &
\colhead{$\chi^2$/d.o.f.} \\
&   $10^{20}$~cm$^{-2}$ & eV & eV  & & eV & erg cm$^{-2}$ s$^{-1}$ &
}
\startdata
 bb & $ 3.65_{-0.13}^{+0.16}  $ &    $101.4_{-0.6}^{+0.5} $  & & &
 & $5.16 \times 10^{-12}$
& 1.36 \\
 bb+abs. edge  &  $3.60_{-0.16}^{+0.21}$ &
 $104.0_{-0.7}^{+0.6}$ & 
$694_{-11}^{+5}$ & 
$0.25_{-0.03}^{+0.03}
 $ & &$5.07 \times 10^{-12}$ & 1.17
\\
 bb+ gauss. line  &  $3.74_{-0.10}^{+0.14}
 $& $102.1_{-0.3}^{+0.5}
 $ & $
754_{-9}^{+8}
 $ &  $4.8_{-0.5}^{+1.0}
 $ & $ 27_{-8}^{+15}
$ & $5.20 \times 10^{-12}$  & 1.20
\\
\enddata
\tablenotetext{a}{All fits have been obtained by fitting
simultaneously data from EPIC PN, MOS1 and MOS2; the absorption model is 
{\it TBabs\/} in XSPEC. The reported
errors are the 68\% ($1 \sigma$) confidence range. 
The number of degrees 
of freedom (d.o.f.) is: 286 (bb), 284 (bb+abs. edge) and 
283 (bb+gauss. line), in the three cases respectively.}
\tablenotetext{b}{Unabsorbed flux measured with EPIC-PN in the
$(0.2 - 2)$ keV band.}
\end{deluxetable}

\begin{deluxetable}{cccc}
\tablecolumns{4} \tablewidth{0pc} \tablecaption{Model Fit Parameters: 
atmospheric models\tablenotemark{a} 
\label{tab2}} \tablehead{ \colhead{$B$} & \colhead{$n_H$} &
\colhead{$kT^\infty$} &
\colhead{$\chi^2$/d.o.f.} \\
 $10^{12}$ G  &  $10^{20}$~cm$^{-2}$ & eV  &  
}
\startdata
 0 & $8.4_{-0.2}^{+0.3}$ & $ 29.8_{-0.3}^{+0.3}$ &
1.94   \\
 1 & $8.3_{-0.2}^{+0.2}$  & $47.3_{-0.4}^{+0.5}$ &  1.99   \\
 10 & $8.6_{-0.2}^{+0.1}$  & $50.0_{-0.4}^{+0.4}$ &  2.19    \\
\enddata
\tablenotetext{a}{All fits have been obtained by fitting
simultaneously data from EPIC PN, MOS1 and MOS2; the absorption model is
{\it TBabs\/} in XSPEC. The reported
errors are the 68\% ($1 \sigma$) confidence range, and the number of 
degrees of freedom (d.o.f.) is 286 in all the three cases.  
The star mass and
radius have been fixed at $1.4$ M$_\odot$ and 10 km, respectively.
The temperature measured by a distant observer, $T^\infty$, and
the effective temperature obtained by the model fit, $T_{eff}$,
are related by $T^\infty = (1 + z)^{-1} T_{eff}$, where $(1 + z)^{-1} = 
[1-2.95(M/M_\odot)/(R/{\rm 10\, km})]^{1/2}$ is the 
gravitational redshift
factor.}
\end{deluxetable}

\clearpage

%%%%%%%%%%%%%%%%%%%%%%%%%%%%%%%%%%%%%%%%%%%%%%%%%%%%%%%%%%%%%%%%%%%%

\end{document}